# A Landauer Limit for Robotic Manipulation

Henry Hess

*Department of Biomedical Engineering, Columbia University, USA*

hhess@columbia.edu

ABSTRACT. A major role of robots is to assist in assembly by moving building blocks and by exerting forces, e.g. to snap parts together. At the molecular scale, diffusive transport and thermal forces permit self-assembly, and molecular robots can only accelerate the process by performing work. This raises the question if – similar to the Landauer principle in computing – there is a lower limit to the work done by a robot for a given acceleration of an assembly process. Here, a brief analysis suggests that a doubling of a reaction rate by robotic manipulation requires at least $k_B T\ln(2)$ in energy expenditure, either to perform mechanical work or to erase information.

Landauer deduced that in order to satisfy the second law of thermodynamics, the erasure of one bit requires an amount of energy equal to at least $k_B T\ln(2)$.[1,2] This insight about a basic process in computing has informed the discussion about the fundamental physical limits of computers[3] as Carnot's limit and its refinements have informed the discussion of the fundamental physical limits of heat engines.[4,5] In the long history of computing devices, energy losses have exceeded the Landauer limit by many orders of magnitude, and only recently has Landauer's limit achieved practical relevance in nanotechnology[6] and also in the study of molecular biology[7].

Here, I aim for a corollary to Landauer's principle related to the physical limits of robotic manipulation. Industrial robots were initially machines to manipulate parts, that is machines to pick things up and put them down in new locations (Fig. 1). Everyday experience teaches that macroscopic things do not move on their own, and significant energy expenditure is required to overcome friction and bring parts into contact. Robotic manipulation has advanced towards the manipulation of smaller and smaller parts,[8] and the first molecular robots have been introduced.[9-11] At the molecular scale, thermal fluctuations enable parts to move, meet, and interact, and robotic manipulation merely accelerates and controls these processes rather than enabling them. At the same time, energy expenditures required to obtain information (e.g. about the position of parts) become significant relative to energy expenditures required for translation, which can themselves drop far below the thermal energy. In analogy to Landauer's calculation of the minimal energy required to erase a bit, can we determine the minimal energy required to e.g. double the rate at which two parts are brought into contact?

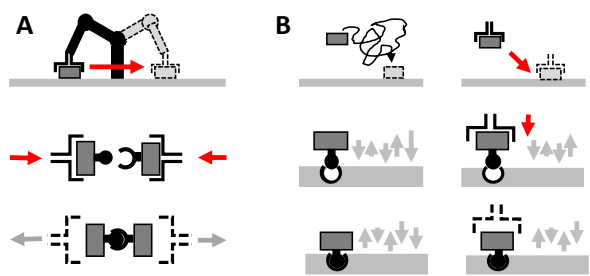

**Figure 1:** A – Macroscale robotic manipulation is transporting parts and can exert forces. B – Brownian motion and thermal forces permit self-assembly at the microscale, and robots can merely accelerate the assembly process.

Following Szilard's approach to studying Maxwell's demon,[12,13] let us consider two particles in a box. Brownian motion will bring the two particles into contact at a rate dependent on their interaction radii, diffusion constants and initial separation; this rate is referred to as the diffusion-limited reaction rate. Subsequently, two interacting particles can react, that is utilize thermal energy to overcome energetic barriers to bond formation at a rate frequently described by the Arrhenius equation.[14] The point of robotic manipulation can be (1) to accelerate the intrinsic rate of contact, that is to bring the molecules or bricks together at a rate considered sufficiently high, and (2) to assist two interacting particles to successfully form a bond, or both. Both objectives will be discussed in succession.

**Accelerating the rate of contact:**

*A basic approach* to accelerating the rate of contact between two diffusing particles is to do work to reduce the volume of the container. The necessary work is composed of an irreversible portion needed to overcome friction and a reversible portion needed to perform the isothermal compression. In this approach, information is not manipulated since the robot merely compresses the container and expands it after the reaction has completed. Because the reaction has reduced the number of particles by one, the work needed to compress the educts exceeds the work gained by isothermal expanding the container with the product by $\Delta W = k_B T \ln(V_f/V_i)$. Since the rate of a diffusion-limited reaction of two particles increases in inverse proportion to the volume of the container, each doubling of the rate of encounters requires at least an amount of work equal to $k_B T \ln(2)$. This minimal energy expenditure is further increased by the irreversible portion of the performed work.

*A robot*, that is a machine capable of sensing, computing and actuating,[15] can pursue a more sophisticated approach: It can (i) fix the locations of the two particles with a certain accuracy, (ii) determine these locations, compute a path from one to the other, and move the particles so that they occupy the same location, (iii) wait for the reaction to proceed, release the product particle and extract work from this isothermal expansion, and (vi) complete the process by erasing the memory where the initial location of the particles is stored (Fig. 2). The energetic costs and gains associated with this generic sequence of grabbing the particles, bringing them into contact, and releasing the product particle are now enumerated:

(i) The position of the two particles is fixed in the fashion of a Maxwell demon by inserting n partitions into the container, which systematically subdivide the space in the most efficient manner, that is by creating subcontainers of equal volume. Following Szilard, it is assumed that this can be done instantaneously and without energetic cost.

(ii) The robot then writes the positions of the two particles into a memory, requiring n bits per particle (one bit per partition), computes a path, and moves the two particles so that they occupy the same subcontainer. The directed movement carries an irreversible energetic cost depending on the drag on the particle, the distance to be covered and the time allocated.

(iii) The reaction proceeds at a rate which is increased in inverse proportion to the volume reduction. The isothermal expansion of the volume containing the product particle yields equilibrium work equal to $W = k_B T \ln(V_f/V_i)$, which is again reduced by an irreversible component.

(iv) The memory contains the initial positions of the two particles, specified by numbering all the subvolumes and writing the position of each particle using $\log_2(V_i/V_f)$ bits. Since erase of each bit requires a minimal energy expenditure of $k_B T \ln(2)$, the total energy required to erase the stored positions is at least $2 k_B T \ln(V_i/V_f)$.

The energy balance of the four steps is an energy expenditure equal to $k_B T \ln(V_i/V_f)$ plus the irreversible work expended in moving the particle and conducting the computation. The difference to the simple compression of the volume discussed in section (a) is that now the energy is expended on manipulating information rather than doing mechanical work, although the minimal amount of energy required to double the rate of encounters requires again at least an amount of work equal to $k_B T \ln(2)$.

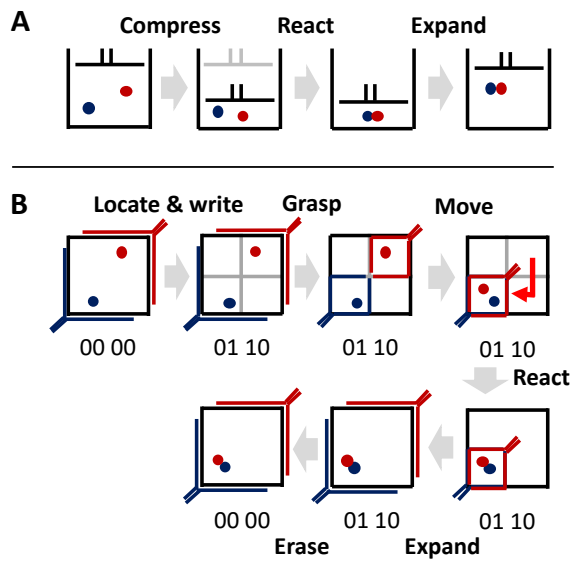

**Figure 2:** A – The frequency of collisions between Brownian particles can be increased by compression, but if the particles are sticky the work expended for compression exceeds the work retrieved during expansion. B – An ideal robot can sense the position of the particles, grasp them (confine them into smaller volumes), compute a path and bring the particles into the same subcompartment, let the reaction occur, obtain work from expanding the subvolume, and expend work to erase the particle positions.

**Increasing the sticking probability after contact:**

Transition state rate theory explains that the sticking probability S in chemical reactions can be described by an Arrhenius-type equation: $S = A \times exp\left(-\frac{E_a}{k_B T}\right)$ where A is a prefactor and $E_a$ is the activation energy, because thermal energy is required to overcome energy barriers to the formation of a bond.[14] The advances in mechanochemistry over the past decades have demonstrated that external forces can be used to alter the activation energy and accelerate or decelerate the reaction.[16] In the case of a constant applied force F, this concept is captured by the Bell equation $S = A \times exp\left(-\frac{E_a + Fx^*}{k_B T}\right)$ where F is positive when the force points in the direction of increasing distance between the particles, and x* is the distance to the transition state.[17, 18] A robot encountering two interacting particles (e.g. a molecule physisorbed to a nanoparticle) can now apply force to accelerate the bond formation. A "smart" way to apply force is to push the particles together only until they reach the transition state, since after crossing the transition state the continuation of the reaction is energetically downhill. In that case, the external work W is exclusively used to access the transition state and we obtain an increased sticking probability of $S(W) = S_0 \times exp\left(\frac{W}{k_B T}\right)$. Again doubling the reaction rate by doubling the sticking probability requires at least an amount of work equal to $k_B T \ln(2)$. Of course, friction during the motion, force applied beyond the transition state, and other losses can increase the actual energy consumption by many orders of magnitude.

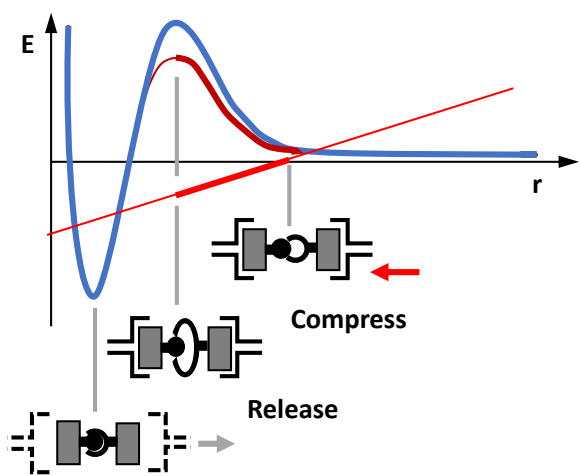

**Figure 3:** If a robot applies a force assisting bond formation, it reduces the activation energy and accelerates the reaction. The robot minimizes its work expenditure if the force is withdrawn immediately after the transition state is passed.

Whichever one of the two objectives a robot pursues to accelerate a reaction, the energetic cost for doubling the reaction rate appears to be at least $k_B T \ln(2)$, expended to do mechanical work or erase information.

One alternative is to heat the system and accelerate self-assembly. A given amount of work can be used to drive a heat pump to warm the system up, let the reaction occur, and recover work while cooling the system. The theoretically possible increase in the rate of collisions per net work expenditure approaches the performance of a robot. However, heating is much more effective in overcoming an energy barrier, since it disproportionally increases the frequency of high energy attempts. A second alternative is to employ a catalyst, which accelerates the formation of bonds, but incurs the penalty of accelerating the reverse process.[19]

Just as in computing, we are often willing to pay a large energetic cost to more rapidly execute certain assembly processes, such as the capture of analyte molecules and their deposition at a sensing site, and in these circumstances molecular robots can find applications even if they are operating far from their thermodynamic limits.[20] However, as the above discussion aims to demonstrate, there are performance limits even if we develop robots which sense and compute at the thermodynamic limit and actuate with negligible friction and other losses.

ACKNOWLEDGMENT

Financial support from the US Army Research Office under W911NF-17-1-0107 and NSF grant CMMI-1662329 is gratefully acknowledged.